\documentclass[lettersize,journal]{IEEEtran}
\usepackage{amsmath,amsfonts}
\usepackage{algorithm}
\usepackage{array}
\usepackage[caption=false,font=normalsize,labelfont=sf,textfont=sf]{subfig}
\usepackage{textcomp}
\usepackage{bm}
\usepackage{stfloats}
\usepackage{url}
\usepackage{color}
\usepackage{xcolor}
\usepackage{verbatim}
\usepackage{graphicx}
\usepackage{multirow}
\usepackage{multicol}
\usepackage{booktabs}
\usepackage{makecell}

\usepackage{algpseudocode}

\begin{document}

\title{Deep Audio-Visual Singing Voice Transcription based on Self-Supervised Learning Models}

\author{Xiangming Gu, Wei Zeng, Jianan Zhang, Longshen Ou, Ye Wang
\thanks{Xiangming Gu, Wei Zeng, Jianan Zhang, Longshen Ou, Ye Wang are with School of Computing, National University of Singapore, Singapore. Ye Wang is the corresponding author.}
\thanks{Manuscript received xxx; revised xxx.}}

\markboth{Journal of \LaTeX\ Class Files,~Vol.~xxx, No.~xxx, xxx~20xx}%
{Shell \MakeLowercase{\textit{et al.}}: A Sample Article Using IEEEtran.cls for IEEE Journals}


\maketitle

\begin{abstract}
Singing voice transcription converts recorded singing audio to musical notation. Sound contamination (such as accompaniment) and lack of annotated data make singing voice transcription an extremely difficult task. We take two approaches to tackle the above challenges: 1) introducing multimodal learning for singing voice transcription together with a new multimodal singing dataset, N20EMv2, enhancing noise robustness by utilizing video information (lip movements to predict the onset/offset of notes), and 2) adapting self-supervised learning models from the speech domain to the singing voice transcription task, significantly reducing annotated data requirements while preserving pretrained features. We build a self-supervised learning based audio-only singing voice transcription system, which not only outperforms current state-of-the-art technologies as a strong baseline, but also generalizes well to out-of-domain singing data. We then develop a self-supervised learning based video-only singing voice transcription system that detects note onsets and offsets with an accuracy of about 80\%. Finally, based on the powerful acoustic and visual representations extracted by the above two systems as well as the feature fusion design, we create an audio-visual singing voice transcription system that improves the noise robustness significantly under different acoustic environments compared to the audio-only systems.

\end{abstract}

\section{Introduction}

\IEEEPARstart{S}{inging} voice transcription (SVT), a task that automatically transcribes note events from singing audio signals, converts our singing voice to music notation. In addition to frame-level pitch estimation \cite{de2002yin, mauch2014pyin, mauch2015computer, kim2018crepe}, which focuses on predicting pitch contour / fundamental frequency (F0) at each time step, SVT pays attention to note events. Each note event includes onset time, offset time, and pitch (in hertz or semitone). This setup makes SVT more challenging but fits many music-related downstream applications. For example, the note events can be applied to music education \cite{yang2022multi}, music therapy \cite{tam2007movement}, human-computer interaction \cite{muller2019computational}, and singing voice synthesis (SVS) \cite{liu2022diffsinger, huang2022singgan}.

In this work, we aim to tackle two main challenges in the SVT that hamper its development:

\textbf{Noise Robustness.} In real-world applications, recorded singing data is often contaminated by musical accompaniment, applause, cheers, and other noises. Previously SVT models, e.g. \cite{mauch2015computer, fu2019hierarchical, hsu2021vocano, wang2021preparation, kum2022pseudo}, are all based only on audio signals, which result in their inability to deal with low signal-to-noise ratio (SNR) scenarios. Among them, only \cite{kum2022pseudo} proposed training an SVT system directly on polyphonic music with both singing and accompaniment. However, this method can not generalize to different SNR levels or noise types (e.g. babble/white/natural noise). Therefore, the solutions to building robust SVT systems still need to be explored. 

\textbf{Label Insufficiency.} Existing annotated SVT datasets are all small-scale, primarily because manual annotation of SVT is a highly time-consuming and demanding task. For instance, the largest manually annotated SVT dataset, MIR-ST500 \cite{wang2021preparation} has about 30 hours, 
much less than 960 hours of speech recordings in LibriSpeech \cite{panayotov2015librispeech} (a benchmark dataset for automatic speech recognition). Several attempts have been made to mitigate the problem of insufficient labeled data \cite{hsu2021vocano, kum2022pseudo, wang2022musicyolo}. Specifically, \cite{hsu2021vocano} separated SVT into onset/offset detection and pitch estimation. For onset/offset detection, they adopted virtual adversarial training (VAT) \cite{miyato2018virtual} to take advantage of unlabeled singing data; for pitch estimation, they directly used an existing model named PatchCNN \cite{su2018vocal}. The separated design made \cite{hsu2021vocano} not an end-to-end framework.  Another work \cite{kum2022pseudo} exploited the benefits from frame-level pseudo labels. Using an existing F0 predictor, \cite{kum2022pseudo} firstly estimated frame-level pitch from singing data without note-level annotations. The estimated continuous pitch is quantized into semitone and then smoothed by median filters. After obtaining the pseudo labels, \cite{kum2022pseudo} followed the noisy student framework \cite{xie2020self} to train the SVT system. However, the evaluation of \cite{kum2022pseudo} was conducted on Chinese singing only. Its generalization to other genres is unexplored. Besides, the unlabeled singing data in \cite{hsu2021vocano} and singing data with frame-level annotations in \cite{kum2022pseudo} are tiny compared to rich speech data. From another perspective, \cite{wang2022musicyolo} formalized onset/offset detection as an object detection task and used YOLOX \cite{ge2021yolox}, which is pretrained on image data. However, the knowledge learned from images is difficult to transfer to singing data. For example, pitch estimation cannot be benefited from the pretrained YOLOX. To summarize, the problem of label insufficiency is still not well resolved.

To improve the noise robustness, we propose introducing multimodal learning by building an audio-visual system for the SVT task. The idea is motivated by both theoretical guarantees \cite{huang2021makes} as well as widespread empirical success of multimodal learning, such as audio-visual speech recognition \cite{afouras2018avsr:deepe2e, ma2021avsr:conformer, shi2022avsr:avhubert, shi2022avsr:robust}, audio-visual active speaker detection \cite{tao2021someone}, etc. Additionally, in the singing domain, \cite{li2021audiovisual, montesinos2021cappella} built audio-visual singing source separation systems. \cite{gu2022mm} built the first multimodal automatic lyric transcription system, MM-ALT, demonstrating that it is robust to the sound contamination, like musical accompaniment. Through our experiments, we observe that video modality can improve overall performance in onset/offset detection and pitch estimation. For example, videos of lip movements can successfully discriminate the transitions between contiguous notes, thus providing extra cues for onset/offset detection. Furthermore, compared to audio-only systems, our audio-visual SVT system demonstrates higher noise robustness in different acoustic environments, including the musical accompaniment, white noise, babble noise, and natural noise.

 To deal with the label insufficiency, we propose adopting self-supervised learning (SSL) models in the SVT task. Previous research has demonstrated that SSL models perform well in downstream tasks after finetuning, even in low-resource scenarios \cite{baevski2020wav2vec, hsu2021hubert, shi2022avsr:avhubert, baevski2022data2vec}. However, there are two obstacles in following this line of work. Firstly, there are no SSL models in the singing domain. To solve this, we choose SSL models trained on rich unlabeled speech data, considering the similarities between speech and singing data. \cite{ou2022towards} has shown that wav2vec 2.0 can be adapted from the speech domain to the ALT task. The adaptation of wav2vec 2.0 benefits from the fact that ALT and ASR are counterpart tasks, which means that the input-output pairs are essentially the same. However, in the scenario of SVT, the labels are the onset/offset/pitch scores instead of the texts in the ASR/ALT task. We must consider the task difference and domain shift between speech and singing, which is our second obstacle. To deal with this, we propose a new strategy for adapting SSL models from the speech domain to the SVT task inspired by \cite{kumar2022fine}. In detail, we assume that the finetuning may distort the pretrained features. Therefore, we propose skipping the finetuning stage of SSL models on downstream tasks in the speech domain and finetune the SSL models directly on the SVT task using linear-probing and full-finetuning (LP-FT). The resulting SVT systems demonstrate high performance on both in-domain (ID) distribution and out-of-domain (OOD) singing data. We summarize the main contributions of this work here:

\begin{itemize}
    \item We propose a new pipeline to adapt the SSL models from the speech domain to the SVT tasks without distortion of pretrained features. Our audio-only SVT system outperforms state-of-the-art technologies on multiple benchmark SVT datasets significantly. It also generalizes to out-of-domain singing data with different languages and styles.
    \item We initialize the task of video-only SVT and demonstrate that videos of lip movements can detect the onsets/offsets of note events through both quantitative results and qualitative analysis.
    \item We curate the first multimodal SVT dataset: N20EMv2. Our audio-visual SVT system shows higher robustness to different types of noise perturbations compared to audio-only SVT systems.
\end{itemize}

\begin{figure}[t]
\centering
\includegraphics[width=\linewidth]{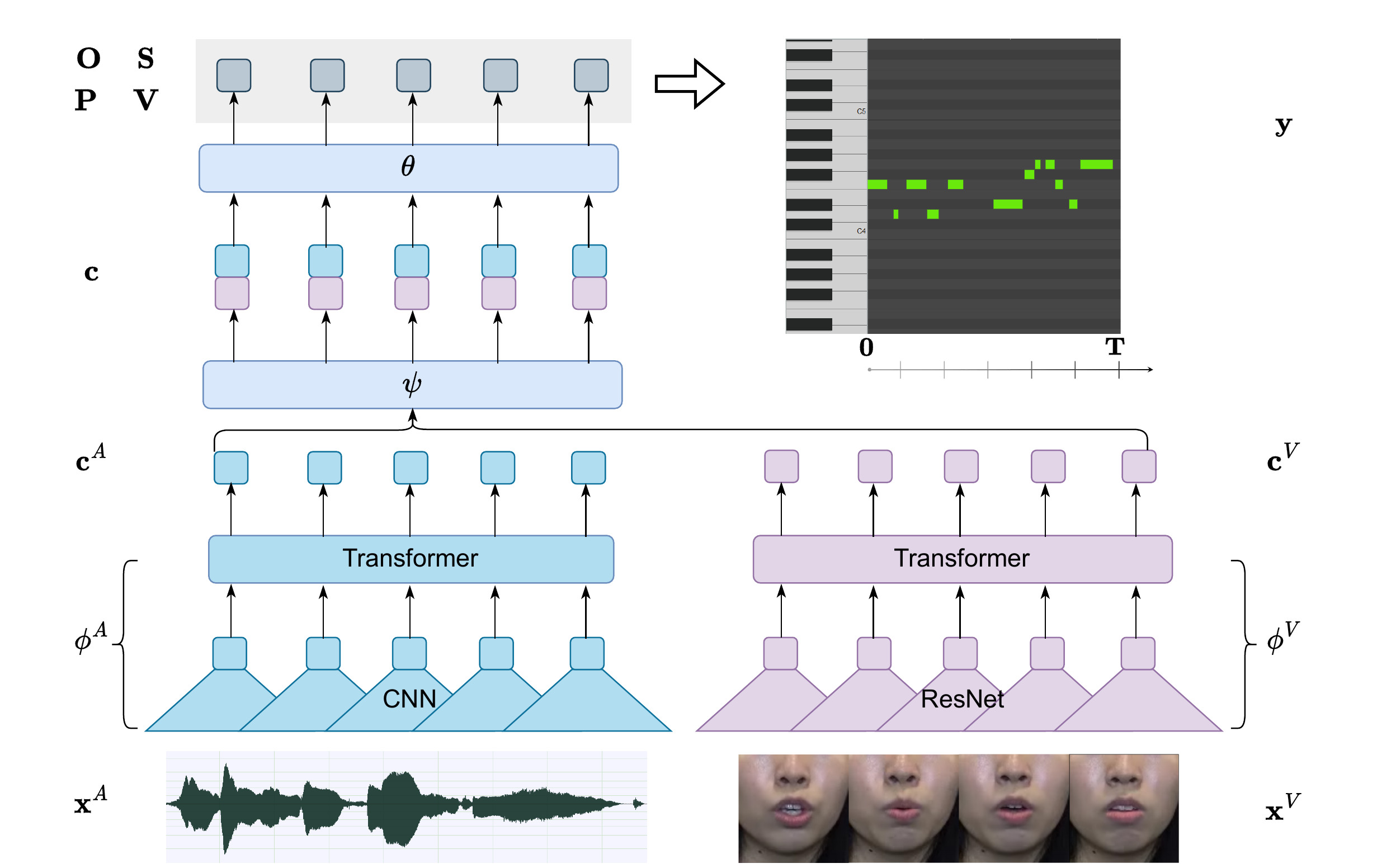}
\caption{Problem setting of the audio-visual SVT.}
\label{fig_SVT}
\end{figure}

\section{Methodology}

\subsection{Problem Setting of SVT}
We first consider the conventional problem setting of audio-only SVT (A-SVT). Suppose the input waveform is $\bm{x}^A$ whose duration is $L$, and the output note events can be represented as $\bm{y}$. Concretely, $\bm{y}$ contains $N$ note events: $\bm{y}=[(o_1, f_1, p_1), ..., (o_n, f_n, p_n),...,(o_N, f_N, p_N)]$, where $o_n$ and $f_n$ are the onset/offset time of $n$-th note, $0\leq o_1<f_1\leq o_2<f_2\leq ...\leq o_N<f_N\leq L$, and $p_n$ is the pitch value. A-SVT task aims to train a neural network to map from $\bm{x}^A$ to $\bm{y}$. We model the training of the A-SVT system as a frame-level classification problem. Firstly, the raw waveform $\bm{x}^A$, which is a 1-D tensor, is fed into the audio-specific feature encoder $\phi^A$ to extract the deep acoustic representations $\bm{c}^A\in \mathbb{R}^{T\times D}$, where $T$ is the number of frames (i.e. time steps), and $D$ refers to the number of dimensions. Since the duration of $\bm{x}$ is $L$, the frame length or frame resolution of acoustic representations is $\frac{L}{T}$. Afterward, each frame of features is sent to a classifier $\theta$ to obtain the frame-level predictions for note events. To supervise the training of $\phi^A$ and $\theta$, we transform the note-level annotations into frame-level annotations. 

Following \cite{wang2021preparation}, the annotation for each frame has four targets, including onset, silence, pitch name, and octave. Since it is difficult to predict offset directly, our model predicts silence instead. Then the offset times $f_1, f_2,...,f_N$ are located at the beginnings of silence.  Onset labels $\bm{O}$ and silence labels $\bm{S}$ are 1-D tensors: $\bm{O}, \bm{S}\in\mathbb{R}^T$. The frames covering the onset times $o_1, o_2,...,o_N$ are marked as onset frames and labeled as 1 (otherwise labeled as 0). Similarly, the frames covering the silent times (no notes) are marked as the silence frames and labeled as 1 (otherwise labeled as 0). Conventionally, the pitch values are labeled as MIDI note numbers from C2 (MIDI number 36, 65.41 Hz) to B5 (MIDI number 83, 987.77 Hz)\footnote{Two adjacent MIDI numbers are differed by one semitone.}. We further split each pitch into a pair of values: octave and pitch name. The octave range is from 2 to 5, and the pitch name range is from C to B, representing 12 notes in each octave. For example, the octave of C3 (MIDI number 48) is three, while its pitch name is C. We also add an octave class and a pitch name class to represent the silence. Therefore, the labels for pitch name and octave are $\bm{P}\in\mathbb{R}^{T\times13}, \bm{V}\in\mathbb{R}^{T\times5}$. The frame-level predictions are concatenated and transformed back into the note events through post-processing, which will be elaborated on later. 

Furthermore, we extend the type of input modality and propose the problem settings of video-only SVT (V-SVT) and audio-visual SVT (AV-SVT). For V-SVT, the difference is replacing the waveform $\bm{x}^A$ with the videos of lip movements $\bm{x}^V$, and the audio-specific feature encoder $\phi^A$ with the video-specific feature encoder $\phi^V$. While for AV-SVT, both audio and video modality are enabled (shown in Fig. \ref{fig_SVT}). We first adopt modality-specific feature encoder $\phi^A$, $\phi^V$ to obtain the features for each modality $\bm{c}^A$, $\bm{c}^V$. Subsequently, another module $\psi$ is used to fuse the features: $\bm{c}=\psi(\bm{c}^A, \bm{c}^V)$.

\subsection{Single-Modal SVT System}
\subsubsection{A-SVT System}
\begin{figure}[t]
\centering
\includegraphics[width=\linewidth]{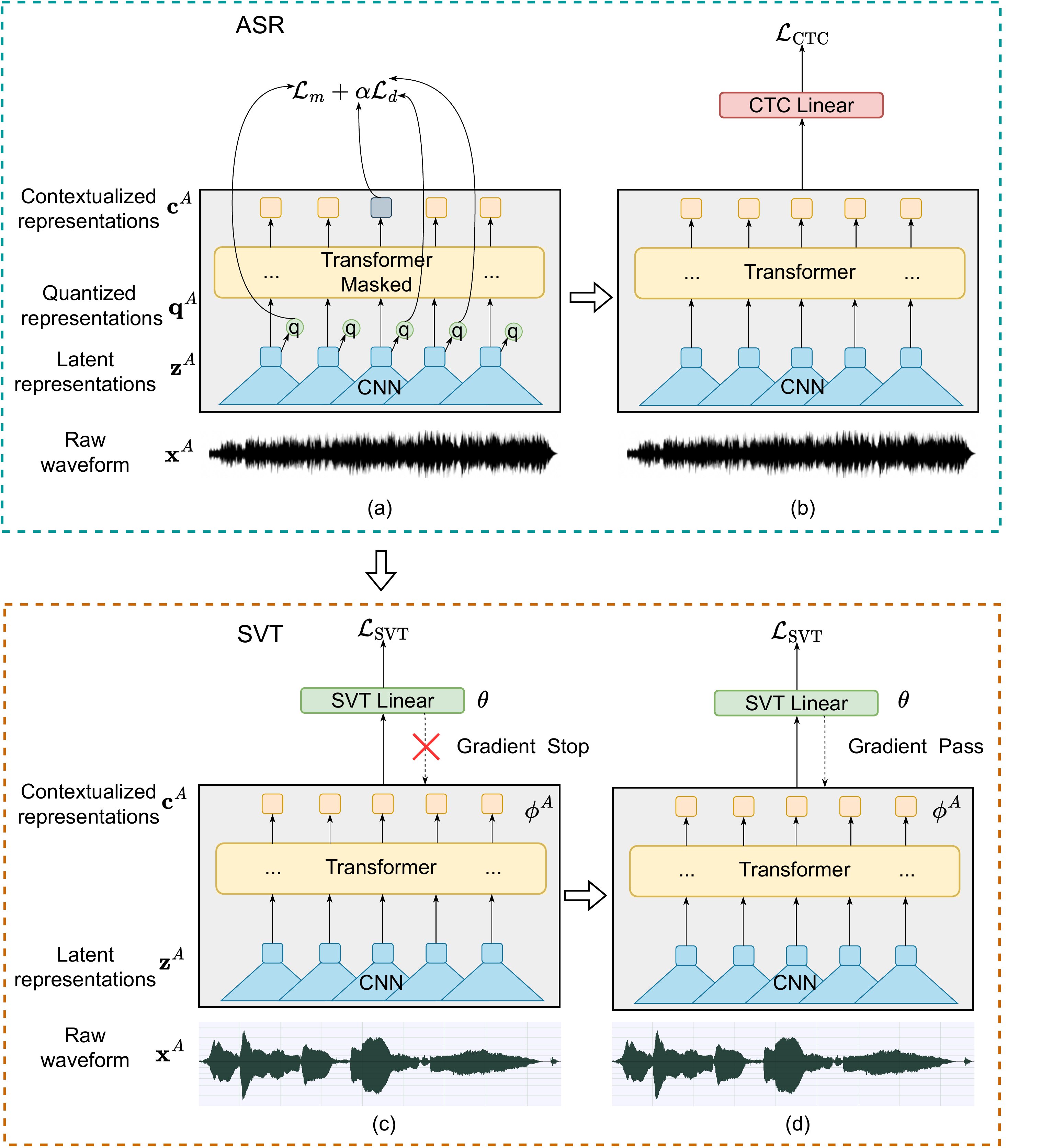}
\caption{Adaptation procedure of wav2vec 2.0 from the speech domain to the SVT task. (a) Pretraining stage of wav2vec 2.0 with unsupervised training objectives; (b) Finetuning stage of wav2vec 2.0 on downstream ASR task; (c) Linear Probing stage of wav2vec 2.0 on the SVT task; (d) Full Finetuning stage of wav2vec 2.0 on the SVT task.}
\label{fig_lp_ft}
\end{figure}

The model architecture of our A-SVT system is visualized in Fig. \ref{fig_lp_ft}(c) and (d). The audio-specific feature encoder $\phi^A$ is parameterized by wav2vec 2.0 Large \cite{baevski2020wav2vec}, and the classifier $\theta$ is a linear layer. In detail, wav2vec 2.0 Large contains a CNN model to extract latent representations $\bm{z}^A$ and a large Transformer to extract contextualized representations $\bm{c}^A$. The CNN has seven temporal convolution blocks with the kernel sizes of $\{10,3,3,3,3,2,2\}$, strides of $\{5,2,2,2,2,2,2\}$ and 512 channels. The design ensures that the frame length of $\bm{z}^A$ is about 20 ms. Transformer has 24 blocks with model dimension 1,024, inner dimension 4,096 for Feed-forward-network (FFN), and 16 heads for multi-head self-attention (MHSA). The output of wav2vec 2.0 is $\bm{c}^A\in\mathbb{R}^{T\times1024}$. Finally, the classifier $\theta$ is a linear layer. The output dimension is 20, including one dimension for onset prediction, one for offset prediction, five for octave prediction, and 13 for pitch name prediction. 

\subsubsection{V-SVT System}\label{avhubert} The video-specific feature encoder $\phi^V$ is parameterized by AV-HuBERT Large \cite{shi2022avsr:avhubert}. AV-HuBERT includes two branches for audio and video input. In our implementation, we disable the audio branch as we only use AV-HuBERT to extract visual representations $\bm{c}^V$. AV-HuBERT has a hybrid ResNet-Transformer architecture. The videos are first handled by a modified ResNet-18 \cite{shi2022avsr:avhubert}. We set the input to the audio branch as zeros, so the fused features only contain the visual information. Afterward, Transformer, whose architecture is the same as that in wav2vec 2.0 \cite{baevski2020wav2vec}, is adopted to extract contextualized representations $\bm{c}^V$. The V-SVT system has the same classifier design as our A-SVT system.

\subsubsection{Model Training}
Suppose the predicted logits for onset/silence/octave/pitch name are $\hat{\bm{O}}$, $\hat{\bm{S}}$, $\hat{\bm{V}}$, $\hat{\bm{P}}$, respectively. To train the whole SVT system, we adopt the following loss:
\begin{align}
    \mathcal{L}_{\text{SVT}}=\frac{1}{T}\sum_{t=1}^T[&\text{BCE}(\sigma(\hat{O_t}), O_t, w_o)+\text{BCE}(\sigma(\hat{S_t}), S_t, w_s)\nonumber\\
    +&\text{CE}(\hat{V_t}, V_t)+\text{CE}(\hat{P_t}, P_t)],
\end{align}
where $\sigma$ refers to the sigmoid activation function. For onset classification, we set positive weight $w_o$ as $15.0$ to amortize the effects of imbalanced distribution in $\bm{O}$. For silence classification, we set positive weight $w_s$ as 1.0. We propose a masked version of SVT loss to enable batch mode training and handle the samples with uneven duration. Suppose the numbers of frames for all samples in a batch are $T^{1},..., T^{b},..., T^{B}$, where $B$ is the batch size. We pad each sample as well as its frame-level annotations to the duration of $T_{max}=\max\{T^{1},..., T^{b},..., T^{B}\}$ with zeros. Then we construct the mask $\bm{M}\in\mathbb{R}^{B\times T_{max}}$ for each batch. Each element $M_t^{b}=1$ if $t\leq T^{b}$. Otherwise, $M_t^{b}=0$. Then the masked SVT loss in batch mode can be written as:
\begin{align}\label{mask}
    &\mathcal{L}_{\text{SVT}}=\frac{1}{\sum_{b=1}^B\sum_{t=1}^{T_{max}}M_t^b}\sum_{b=1}^B\sum_{t=1}^{T_{max}}M_t^b[\text{BCE}(\sigma(\hat{O_t^b}), O_t^b, w_o)\nonumber\\
    &+\text{BCE}(\sigma(\hat{S_t^b}), S_t^b, w_s)+\text{CE}(\hat{V_t^b}, V_t^b)
    +\text{CE}(\hat{P_t^b}, P_t^b)]
\end{align}

\begin{algorithm}
\caption{Adaptation of the SSL models from the speech domain to the SVT task}\label{alg1}
\begin{algorithmic}
\Require SSL model $\phi^{(0)}$ which has been pretrained under objective $\mathcal{L}_{\text{SSL}}$, randomly initialized classifier $\theta^{(0)}$, learning rates $\gamma_1, \gamma_2$ for $\theta$ and $\phi$, iterations $K_1, K_2$ for linear probing and full finetuning.
\State \textcolor{red}{Skip the stage of finetuning on the ASR task}
\For{$k=1$ \textbf{to} $K_1+K_2$}
\State $\theta^{(k)}=\theta^{(k-1)}-\gamma_1\frac{\partial\mathcal{L}_{\text{SVT}}}{\partial\theta^{(k-1)}}$
\If{$k\leq K_1$}
\State \textcolor{blue}{$\phi^{(k)}=\phi^{(k-1)}$} \Comment{Linear Probing}
\Else
\State \textcolor{blue}{$\phi^{(k)}=\phi^{(k-1)}-\gamma_2\frac{\partial\mathcal{L}_{\text{SVT}}}{\partial\phi^{(k-1)}}$} \Comment{Full Finetuning}
\EndIf
\EndFor
\end{algorithmic}
\end{algorithm}
We propose a new strategy to adapt self-supervised learning (SSL) models from the speech domain to the SVT task. Before delving into our algorithm, we first recap the training of SSL models. Typically, SSL models are firstly pretrained under unsupervised objectives. Afterward, they are finetuned using labeled data pairs on downstream tasks. For wav2vec 2.0, the unsupervised objective is a combination of contrastive loss and diversity loss: $\mathcal{L}_{\text{SSL}}=\mathcal{L}_m+\alpha\mathcal{L}_d$, as shown in Fig. \ref{fig_lp_ft} (a). Specifically, the latent representations $\bm{z}^A$ are also sent to a quantization module to learn discrete units $\bm{q}^A$. The diversity loss $\mathcal{L}_d$ ensures the equal usage of codebook entries of quantization modules. The contrastive loss can be written as:
\begin{equation}
    \mathcal{L}_m=-\log\frac{\exp(sim(c_t^A, q_t^A)/\kappa)}{\sum_{\widetilde{q}^A\in Q_t^A}\exp(sim(c_t^A, \widetilde{q}_t^A)/\kappa)},
\end{equation}
where $\widetilde{q}_t^A\in Q_t^A$ refers to the candidate quantized representations, including one positive and $K$ negatives, $sim$ refers to cosine similarity, and $\kappa$ is a temperature hyper-parameter. During the pretraining, the input to the Transformer will be randomly masked. In \cite{baevski2020wav2vec}, wav2vec 2.0 was then finetuned on the ASR task using Connectionist Temporal Classification (CTC) loss \cite{graves2006e2e:ctc}. For AV-HuBERT, the pretraining stage requires the participation of both audio and video modality. AV-HuBERT alternates feature clustering and masked prediction to perform SSL. The clusters, which are regarded as the target labels for masked prediction, are assigned by clustering audio-visual features. The masked prediction loss is a cross-entropy loss. After pretraining, AV-HuBERT is finetuned on the speech recognition task using CTC loss \cite{graves2006e2e:ctc} and sequence-to-sequence (S2S) \cite{bahdanau2016end} loss. We refer the readers to \cite{baevski2020wav2vec, shi2022avsr:avhubert} for more details. 

Similar to \cite{kumar2022fine}, we find that finetuning on speech recognition tasks distorts the pretrained features of SSL models, thus affecting their performance on SVT tasks in both in-domain (ID) and out-of-domain (OOD) scenarios. Therefore, we skip finetuning the SSL models on speech recognition tasks. We then conduct linear probing on the classifier $\theta$ on SVT tasks before fully finetuning the feature encoder and classifier. We also assume that linear probing mitigates the catastrophic forgetting of SSL models. Furthermore, we adopt smaller learning rates for SSL models than the classifier, similar to \cite{ou2022towards}. The detailed algorithm is described in Alg. \ref{alg1}. Taking wav2vec 2.0 as an example, we visualize this procedure in Fig. \ref{fig_lp_ft}, following the order of $(a)\rightarrow(c)\rightarrow(d)$.

\begin{figure}[t]
\centering
\includegraphics[width=0.8\linewidth]{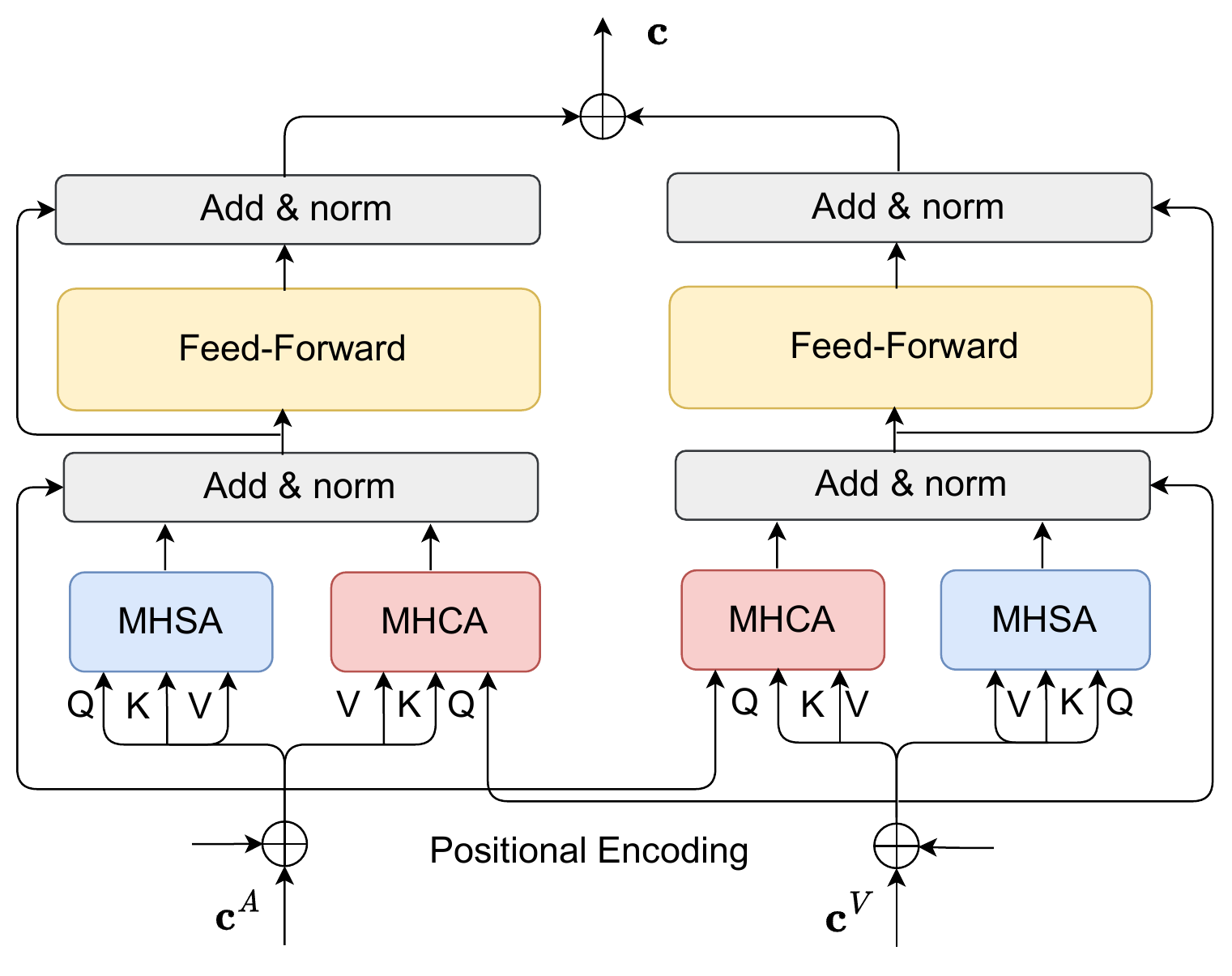}
\caption{Architecture of the Residual Cross Attention (RCA) module.}
\label{fig_rca}
\end{figure}

\subsubsection{Post Processing}
Our post-processing procedure follows \cite{wang2021preparation}. Given the predictions $\hat{\bm{V}}$, $\hat{\bm{P}}$ of the SVT system, we can determine the predicted MIDI number (or silence) of each frame. Afterward, we traverse all frames to search for all note events. For each note event, we first determine its onset time. If onset prediction $\hat{O_t}$ is larger than $0.4$ (onset threshold) and $\hat{O_t}$ is a local maximum, $(t-1)\frac{L}{T}$ is the onset time. Then $(t'-1)\frac{L}{T}$ is the offset time under the condition of $t'=\arg\min (\hat{S_{t'}}>0.5)\,\, \text{and}\,\, t'>t$ ($\hat{S_{t'}}$ is the silence prediction). The MIDI number of this note is the mode of predicted MIDI numbers between $t$-th and $t'$-the frame. It is noticed that the frame resolution $\frac{L}{T}$ makes significant contributions to the accuracy of SVT tasks.

\subsection{Audio-Visual SVT System}\label{av_SVT}

\subsubsection{Modality Fusion}
Compared to the A/V-SVT systems, our AV-SVT system has a feature fusion module $\psi$ to fuse the representations from both audio and video modality. The acoustic representations $\bm{c}^A$ are extracted by wav2vec 2.0 $\phi^A$ while the visual representations $\bm{c}^V$ are extracted by AV-HuBERT $\phi^V$. To align the frame resolution between $\bm{c}^A$ and $\bm{c}^V$, the frame-rate of video input $\bm{x}^V$ is set as 50 Hz instead of 25 Hz in \cite{shi2022avsr:avhubert}. We include an experiment in the supplement to validate that video input with 50 Hz is empirically better than that with 25 Hz. Due to its performance superiority, we parameterize $\psi$ using residual cross attention (RCA) proposed in \cite{gu2022mm}. We visualize the structure of RCA in Fig. \ref{fig_rca}, where MHSA stands for multi-head self-attention while MHCA refers to multi-head cross-attention. The basic idea of RCA is to add cross attention between acoustic features and visual features as shortcuts to augment self-attention.

\subsubsection{Model Training}
Both wav2vec 2.0 and AV-HuBERT are large-scale. Therefore, to mitigate the demanding requirements for GPU memories, we propose to train our AV-SVT system in two stages, similar to \cite{pan2022leveraging}. In the first stage, we train the A-SVT system and V-SVT systems separately, following Alg. \ref{alg1}. Then we use trained audio-specific feature encoder $\phi^A$ and video-specific feature encoder $\phi^V$ in our AV-SVT system. In the second stage, we fix the weights of feature encoders and train the feature fusion module and classifier together.

\section{Datasets}

\subsection{Benchmark A-SVT Datasets}

MIR-ST500 \cite{wang2021preparation} is the largest A-SVT dataset with human annotations. It has 500 Chinese pop songs (about 30 hours), including 400 songs for training and 100 songs for evaluation. TONAS \cite{gomez2013towards} and ISMIR2014 \cite{molina2014evaluation} are two small datasets only to evaluate the A-SVT systems in out-of-domain (OOD) scenarios, considering their different styles, languages, and annotation processes. TONAS has 72 Flamenco songs (36 minutes in total duration), while ISMIR2014 has 14 songs sung by children, 13 by male adults and 11 by female adults (38 pop songs, 19 minutes in total duration). What is worthy of attention is that pitch values of MIR-ST500 and TONAS are annotated as semitones while ISMIR2014 is annotated in cent resolution (1 semitone = 100 cents).

\subsection{N20EMv2 Dataset}
\begin{table}[t]
\caption{Statistics of N20EMv2 train / valid / test set}
\centering
 \begin{tabular}{ l | c | c } 
  \toprule
  Set & Duration & Number of songs \\ 
  \midrule
  Total & 8 h 22 min & 157 \\ 
  Train & 6 h 26 min & 123 \\ 
  Valid & 47 min & 16 \\ 
  Test & 69 min & 18 \\ 
  \bottomrule
 \end{tabular}
 
 \label{tbl-n20emv2}
\end{table}
There are no datasets to support the training and evaluation of video-only / audio-visual SVT systems, so we curate our own dataset N20EMv2\footnote{We will release the N20EMv2 dataset soon.}. It is based on the N20EM dataset \cite{gu2022mm} since N20EM provides synchronized audio and video of singing data. We refer the readers to \cite{gu2022mm} for the details about how N20EM was collected. Here we only highlight the changes of N20EMv2 compared to N20EM. 

Firstly, in N20EMv2, each sample is a whole song instead of an utterance in N20EM. Since train/valid/test sets of the N20EM dataset have no overlapping songs, we keep the same data division in N20EMv2. The statistics of the N20EMv2 dataset are shown in Table \ref{tbl-n20emv2}. The total duration of N20EMv2 is longer than that of N20EM since the silent utterances are removed in N20EM. Secondly, we use videos of lip movements at a frame rate of 50 Hz in N20EMv2 instead of 25 Hz in N20EM. Most importantly, we provide song-level annotations for the A-SVT / V-SVT / AV-SVT tasks in N20EMv2. The label format is the same as MIRS-ST500 \cite{wang2021preparation}. 

\begin{figure}[t]
\centering
\includegraphics[width=\linewidth]{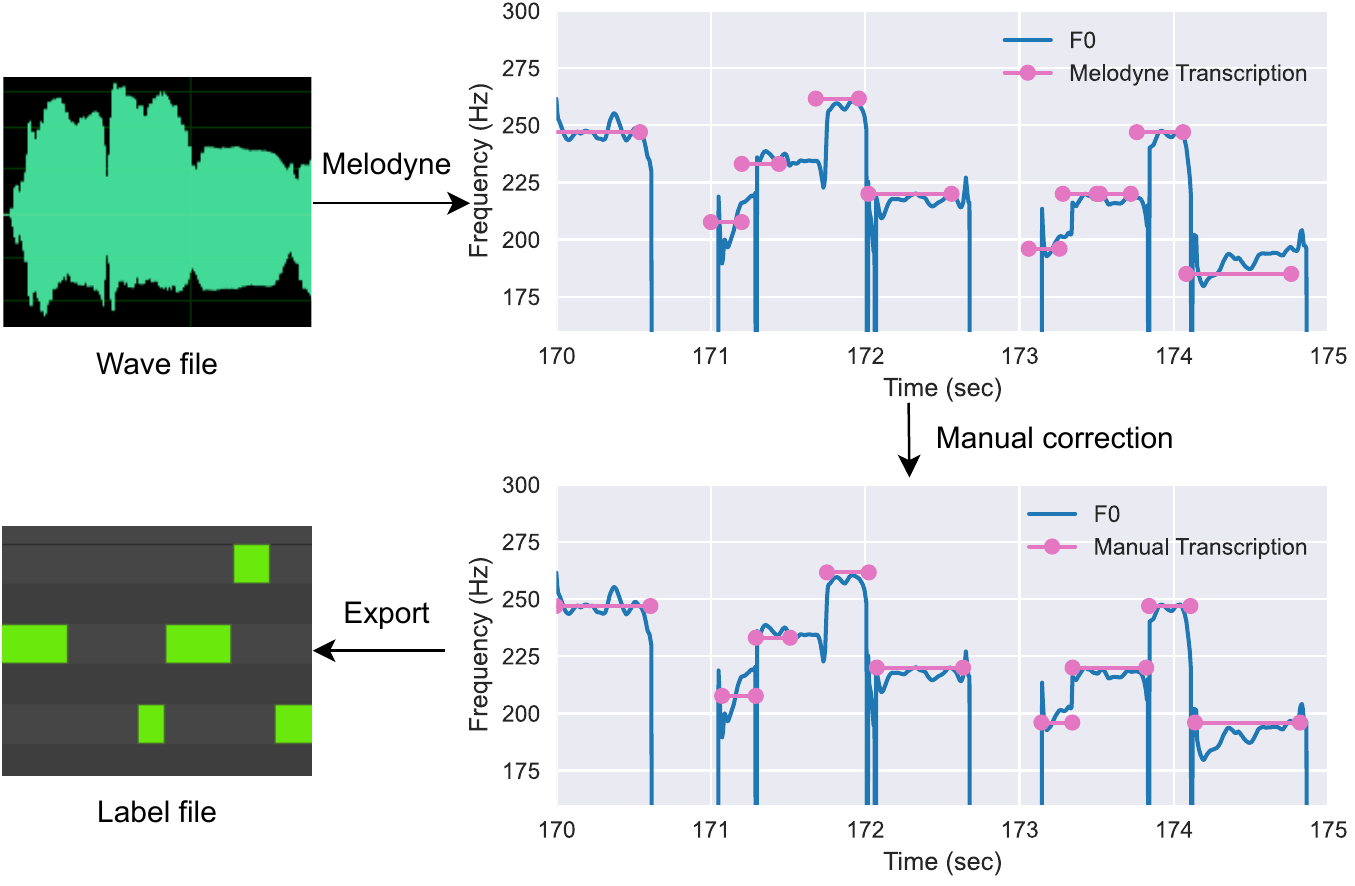}
\caption{Two stages of our annotation process for the N20EMv2 dataset.}
\label{annotation}
\end{figure}

To improve the quality of annotation, we follow a coarse-to-fine manner. The annotation process is shown in Fig. \ref{annotation}. In the first stage, the professional digital signal processing software Melodyne\footnote{\url{https://www.celemony.com/en/melodyne/what-is-melodyne}} is adopted to obtain coarse annotations. Then in the second stage, two experts manually adjust onset/offset/pitch by playing and comparing label and audio tracks simultaneously from an interface comprising spectrogram, waveform, and MIDI notes. In this stage, we set several rules to ensure consistency between different annotators. We include these rules and detailed annotation procedures in the supplement. Our coarse-to-fine annotation procedures ensure that our curated N20EMv2 dataset has higher annotation quality than the MIR-ST500 dataset, which is validated through additional experiments in the supplement. 

\begin{table*}[t]
\caption{COnPOff / COnP / COn F1-score (\%) of different A-SVT systems on MIR-ST500 test set / TONAS / ISMIR2014. We compare our SVT system to state-of-the-art approaches. The best results are marked as \textbf{bold face} while the second-best results are highlighted using \underline{underline}.}
	\centering
	\begin{tabular}{l|l|c|c|c|c|c|c|c}
		\toprule
		Dataset & Metric (\%) $\uparrow$ & Tony \cite{mauch2015computer} & HCN \cite{fu2019hierarchical} & VOCANO \cite{hsu2021vocano} & EfficientNet-b0 \cite{wang2021preparation} & $\text{JDC}_{note}$ \cite{kum2022pseudo} & Ours variant 1 & Ours variant 2 \\
		\midrule
		\multirow{3}{6em}{\textbf{MIR-ST500}} & COnPOff & - & -& -& 45.78 & 42.23 & \underline{52.39} & \textbf{52.84} \\
		& COnP & - & -& -& 66.63 & 69.74 & \textbf{70.73}& \underline{70.00} \\
		& COn & - & -& -& 75.44 & 76.18 & \textbf{78.32} &\underline{78.05} \\
		\midrule
		\multirow{3}{6em}{\textbf{TONAS}} & COnPOff &  - & -& -& \,\,9.57 & - & \underline{12.71} &\textbf{24.08} \\
		& COnP &  - & -& -& 19.65 & - &\underline{25.24}& \textbf{36.87} \\
		& COn &  - & -& -& 42.41 & - &\underline{52.77}&\textbf{64.38} \\
		\midrule
		\multirow{3}{6em}{\textbf{ISMIR2014}} & COnPOff &  50 & 59.4 & \textbf{68.38} &  49.55 & - & 52.36 &\underline{62.42} \\
		& COnP &  68 & - & \textbf{80.58} & 63.63 & - & 70.38 &\underline{75.91} \\
		& COn &  73 & 79.0 & 84.04 & 79.16 & - & \underline{92.77} &\textbf{93.02} \\
		\bottomrule
	\end{tabular}
    
    \label{tbl-exp-B3}
\end{table*}
\section{Experiments}
In this section, we first conduct experiments on our audio-only SVT system and video-only SVT system to (1) evaluate our model design and adaptation strategy; (2) demonstrate that our models can extract powerful acoustic and visual representations. Then we fuse the above features from audio and video modality and build our audio-visual SVT system. We test its noise robustness in the environments with different noise types and SNR levels.

\subsection{Experimental Setup}
We run our experiments based on the SpeechBrain platform \cite{ravanelli2021speechbrain}\footnote{Our code repo: \url{https://github.com/guxm2021/SVT_SpeechBrain}}. For data pre-processing, we first perform source separation on audio signals to isolate the vocal part using spleeter \cite{hennequin2020spleeter}. To meet the input requirements of wav2vec 2.0 \cite{baevski2020wav2vec}, we down-sample the vocal audio to a 16 kHz sampling rate and convert it to mono-channel if necessary. We simulate noisy environments by mixing the vocal audio with noise according to different SNR, which we will explain later. We follow \cite{shi2022avsr:avhubert, gu2022mm} to process the video signal and perform data augmentation.

Our experiments are conducted on an AMD EPYC 7302P 16-core CPU and two RTX A5000 GPUs. Unless specified otherwise, we choose the following training configurations. For single-modal SVT experiments, we train the model using the Adam optimizer \cite{kingma2014adam} for ten epochs, including two epochs for linear probing and eight epochs for full finetuning. The learning rate for the classifier layer is $3\times10^{-4}$ while the learning rate for the feature encoder is $5\times10^{-5}$. We adopt the Newbob technique to schedule the above learning rates with factors of $0.8$ and $0.9$, similar to \cite{ou2022towards}. The conventional performance metrics of SVT systems include F1-scores of the COnPOff (Correct onset, pitch, and offset), COnP (Correct onset, pitch), and COn (Correct onset). Their definitions and implementations can be found in \cite{raffel2014mir_eval, molina2014evaluation}\footnote{\url{https://github.com/craffel/mir_eval}}. For fair comparisons with previous approaches, the pitch tolerance is set as 50 cents, the onset tolerance is set as 50 ms, and the offset tolerance is set as the maximum of 50 ms and $0.2\times$note duration. In experiments related to the N20EMv2 dataset, we also adopt the F1-score of the COff (Correct offset) metric to evaluate the performance on offset detection.

The metrics we mentioned above are computed on song level. However, loading a whole song to GPU memory is a bottleneck since the duration of each song is about 3-5 minutes, and self-supervised learning (SSL) models are large-scale. To mitigate the demands for GPU memory, we split each song into segments with 5 s (the last segment may last 2.5-7.5 s). In our experiments, the segments are set not to overlap each other. This procedure is directly conducted on the samples and their corresponding frame-level annotations. We set the batch size as 8 for training and 1 for evaluation.

\subsection{A-SVT Experiments}

\subsubsection{Comparison with state-of-the-art technologies}
Our work is the first attempt to adapt self-supervised learning (SSL) models from the speech domain to the SVT tasks. Specifically, the input to our A-SVT system is the raw waveform of singing data instead of manually designed acoustic features, such as constant-Q transform (CQT) in \cite{wang2021preparation}, generalized cepstrum of spectrum (GCoS) in \cite{fu2019hierarchical, hsu2021vocano}, and spectrogram in \cite{kum2022pseudo}. To demonstrate the superiority of this design, we first train our A-SVT system on the MIR-ST500 training set, which is marked as ``Ours variant 1". The results are shown in Table \ref{tbl-exp-B3}. The evaluation on the MIR-ST500 test set can be considered an in-domain (ID) test, while the evaluations on TONAS / ISMIR2014 are out-of-domain (OOD) tests. For ID testing, our A-SVT system outperforms the Efficient-b0 \cite{wang2021preparation} and $\text{JDC}_{note}$ \cite{kum2022pseudo} significantly in terms of COnPOff / COnP / COn. Especially for the metric of COnPOff, our A-SVT system exceeds the previous state-of-the-art (SOTA) performance by a large margin ($6.61\%$ F1-score). For OOD testing, our A-SVT system still performs better than EfficientNet-b0, which indicates the effectiveness of our model architecture design and proposed training strategy. We note that the performances on TONAS are much worse than the MIR-ST500 test set and ISMIR2014. The reason is that TONAS consists of Flamenco songs while other datasets are mostly pop songs, resulting in a large distribution shift.

We also train another A-SVT system on the mixture of MIR-ST500 and N20EMv2 training sets (marked as ``Ours variant 2" in Table \ref{tbl-exp-B3}) to take advantage of our new curated N20EMv2 dataset. Apart from a high performance for ID testing, ``Ours variant 2" demonstrates much better generalization abilities on singing data from unseen domains (OOD test). Specifically, ``Ours variant 2" achieves state-of-the-art performances in terms of COnPOff / COnP / COn on the TONAS dataset and COn on the ISMIR2014 dataset. Also the performance of ``Ours variant 2" is close to state-of-the-art \cite{hsu2021vocano} in terms of COnPOff / COnP on the ISMIR2014 dataset even with quantization errors. In contrast to the MIR-ST500 / TONAS / N20EMv2 datasets, which are annotated in semitones, the pitch values in ISMIR2014 are annotated in cents, which puts our A-SVT system at a disadvantage. Following \cite{wang2021preparation} and modern musical notation, our current design uses a 12-tonal equal temperament system with semitonal resolution, which is more practical in real-world applications. To summarize, wav2vec 2.0 can learn great acoustic representations for the SVT tasks.

\begin{table}[t]
\caption{COnPOff / COnP / COn / COff F1-score (\%) of our A-SVT / V-SVT systems on N20EMv2 valid / test set.}
	\centering
	\begin{tabular}{l|l|c|cc}
		\toprule
		Dataset & Metric & Audio & \multicolumn{2}{c}{Video}\\
        & (\%) $\uparrow$ & Tolerance 1 & Tolerance 1 & Tolerance 2\\
		\midrule
  \multirow{4}{4.5em}{\textbf{N20EMv2 valid}} & COnPOff &  61.83  & 4.45 & 9.27\\
		& COnP & 68.42 & 6.16 & 11.79\\
		& COn & 92.18 & 77.14 & 88.69\\
        & COff & 89.80 & 74.68 & 83.01\\
		\midrule
		\multirow{4}{4.5em}{\textbf{N20EMv2 test}} & COnPOff & 73.06 & 6.84 & 15.25\\
		& COnP & 79.56 & 8.79 & 18.53\\
		& COn & 93.66 & 78.62 & 88.64\\
        & COff & 91.78 & 78.83 & 84.48\\
		\bottomrule
	\end{tabular}
    
    \label{tbl-exp-B4}
\end{table}
\subsubsection{Baseline performance for N20EMv2 dataset}
We use the same A-SVT system trained on the mixture of MIR-ST500 and N20EMv2 training sets (``Ours variant 2") to build the baseline for the N20EMv2 valid/test set. As shown in Table \ref{tbl-exp-B4} (``Tolerance 1" refers to the default onset/offset/pitch tolerance), we observe that the accuracy of predictions on the test set is consistently higher than that on the validation set. Moreover, the performance of onset detection is slightly better than offset detection. One reason could be that the model cannot confidently capture the ending of a note, as notes usually decay rather than cut off suddenly.

\subsection{V-SVT Experiments}
\subsubsection{Baseline performance for N20EMv2 dataset}
We build our V-SVT system based on the adaptation of AV-HuBERT. The curated N20EMv2 is the first dataset for the V-SVT task, so to build the baseline, we train our V-SVT system on the N20EMv2 training set and evaluate its performance on the N20EMv2 valid/test set. The results are summarized in Table \ref{tbl-exp-B4}. We find that using the video of lip movements, our V-SVT system can achieve almost $80\%$ F1-score in terms of onset/offset detection under the default tolerance. This result is noteworthy as it can compete with previous A-SVT systems' performances on the two metrics. Furthermore, we relax the tolerance to investigate the potential of our V-SVT system. In detail, we set the onset tolerance as 100 ms, the offset tolerance as the maximum of 100 ms and $0.2\times$note duration, and the pitch tolerance as 100 cents (labeled as ``Tolerance 2" in Table \ref{tbl-exp-B4}). We notice that the COn F1-score reaches about $89\%$, suggesting that within the range of $\pm 50$ ms, our V-SVT system can accurately detect almost all onsets. For pitch estimation, even with only video, our V-SVT system can hint at the discrimination of different pitches. Therefore, we conclude that AV-HuBERT can learn powerful visual representations for the SVT tasks.

\begin{figure}[t]
\centering
\includegraphics[width=\linewidth]{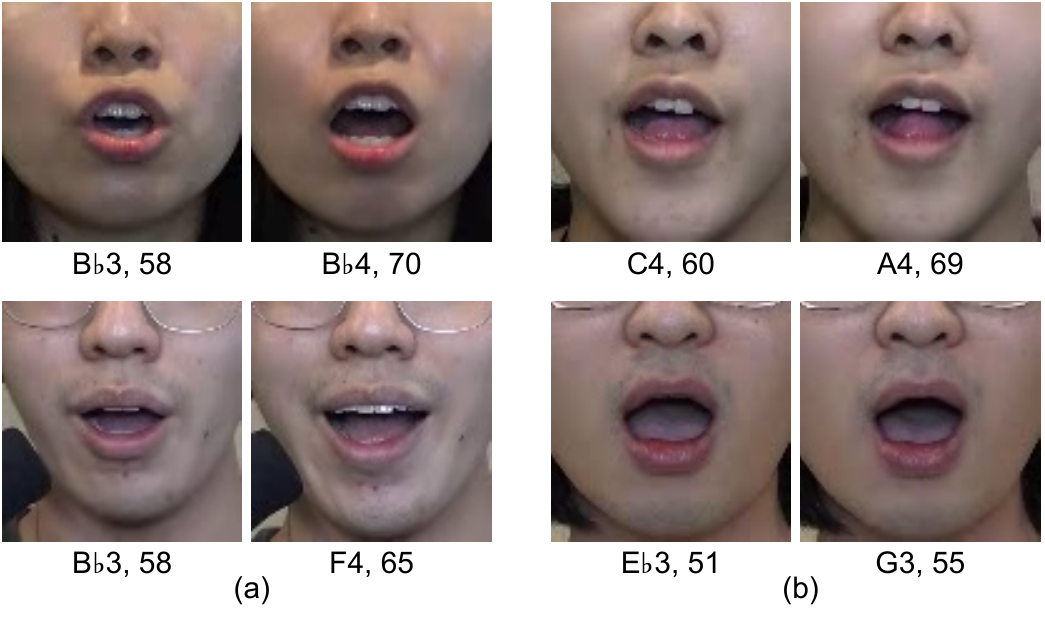}
\caption{Examples of (a) different mouth shapes for the same pronunciation with different pitches and (b) the same mouth shape for the same pronunciation with different pitches.}
\label{video_examples}
\end{figure}

\subsubsection{Analysis of our V-SVT results}
To interpret the high performance of our V-SVT system on onset/offset detection, we assume the reason is that our V-SVT system can detect the transitions of consecutive note events by recognizing small changes in the mouth shape. The pitch of each note event reflects the acoustic information, which is difficult to be captured by video only. The performances of our V-SVT system in COnPOff / COnP demonstrate that it can roughly differentiate between mouth shapes. However, the mouth shapes are not sufficient to predict the pitches. As visualized in Fig. \ref{video_examples} (a), in some cases, different mouth shapes of the same singer correspond to various pitch labels. Our V-SVT system can identify these cases. As shown in Fig. \ref{video_examples} (b), the mouth shapes are the same, but the ground truth MIDI numbers differ. Our V-SVT system will likely fail in these cases.

\begin{figure}[t]
\centering
\includegraphics[width=\linewidth]{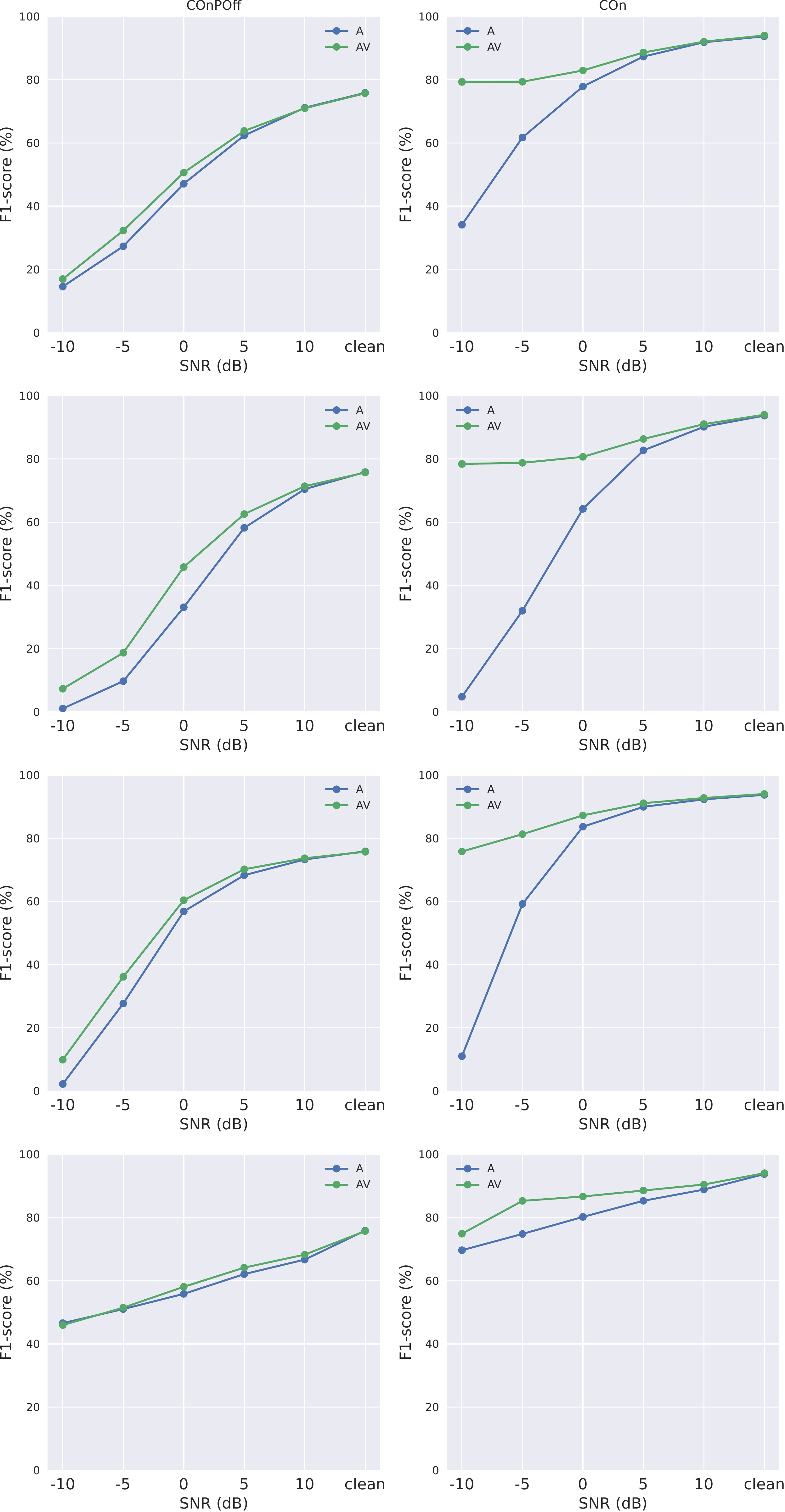}
\caption{Quantitative comparison of our A-SVT system and AV-SVT system on the N20EMv2 test set. We compare the COnPOff, and COn two metrics of two SVT systems under the same noise perturbations for each row. From top to bottom, we use musical accompaniment, babble noise, white noise, and natural noise as the perturbations.}
\label{fig_noise}
\end{figure}

\subsection{AV-SVT Experiments}
To build our AV-SVT system, we follow the training strategy explained in Section \ref{av_SVT}. We use the feature encoders (wav2vec 2.0 and AV-HuBERT) in our best-performing A-SVT and V-SVT systems described above. We then train the feature fusion module and the classifier on the N20EMv2 training set using a larger learning rate with $3\times10^{-3}$ for ten epochs. The motivation behind AV-SVT is taking advantage of video modality to augment the noise robustness under acoustic environments. Therefore, we synthesize noisy audio signals using four different types of noise, including the musical accompaniment, babble noise, white noise, and natural noise\footnote{We include some noisy audio samples in the supplement.}. The babble and natural noise are created based on MUSAN dataset \cite{snyder2015musan}. We set different noise levels (SNR), including -10, -5, 0, 5, 10 dB, and $\infty$ (clean, no noise). We train and evaluate our AV-SVT system under each scenario and report the results in Fig. \ref{fig_noise}. To achieve fair comparisons with our A-SVT system, we follow the same training procedure as our AV-SVT system to train our A-SVT system. To enable the feature fusion module, we set video inputs as zeros.

\subsubsection{Quantitative analysis of AV-SVT system}
In Fig. \ref{fig_noise}, we compare our A-SVT / AV-SVT system on the N20EMv2 test set in terms of the COnPOff and COn metrics. The curves of COnP are similar to that of COnPOff; the curves of COff are similar to that of COn. We visualize the complete comparisons in the supplement. Our results show that the AV-SVT system consistently outperforms the A-SVT system across different noise levels under different noise types. The improvements brought by the video modality are significant in low SNR scenarios. The performance gaps between two SVT systems are narrowed with the increase of SNR since the contributions of the video modality are diluted in less noisy environments. With the assistance of video modality, our AV-SVT system exceeds the A-SVT system by a large margin in COn, which coincides with our assumption. It is undeniable that with the video modality, the overall performance (COnPOff) of the AV-SVT system can also be improved. Compared to the other three noise types, the improvements under natural noise are limited (the last row). As natural noise is short in intervals, we presume that temporary perturbations cause less damage to the A-SVT system than continued perturbations such as the musical accompaniment, babble noise, and white noise. 

\subsubsection{Qualitative analysis of AV-SVT system} 

\begin{figure}[t]
\centering
\includegraphics[width=\linewidth]{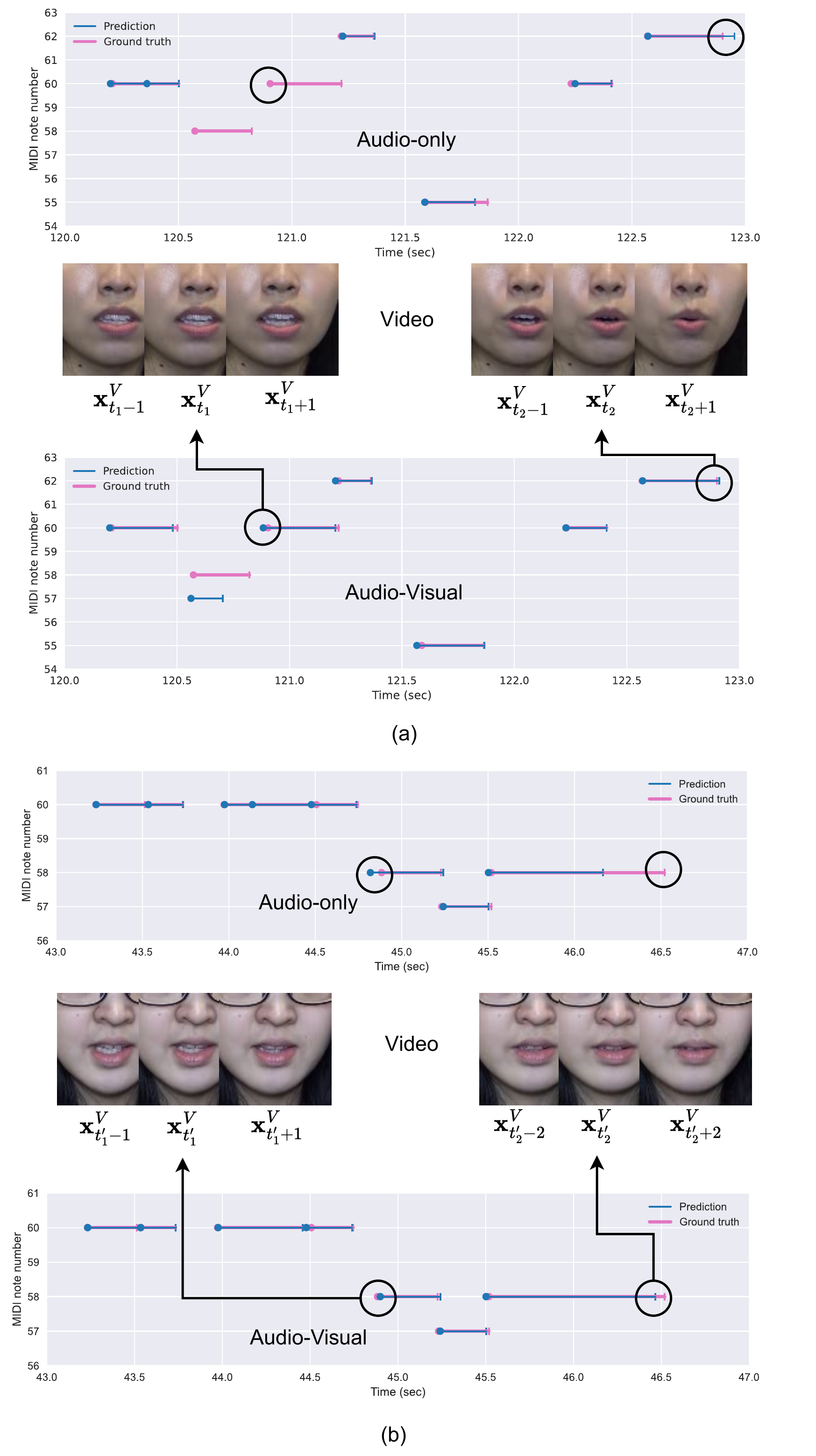}
\caption{Qualitative comparison of our A-SVT and AV-SVT systems on the N20EMv2 test set (a) with 0 dB babble noise perturbation; (b) with 0 dB perturbation of the musical accompaniment.}
\label{fig_qual}
\end{figure}

To further present the effectiveness of our multimodal design, we first visualize the predictions of our A-SVT and AV-SVT systems in an environment of 0 dB babble noise in Fig. \ref{fig_qual}(a). From 120 s to 123 s in the selected song, there are seven notes in the ground truth. Our AV-SVT system predicts seven notes, while our A-SVT system only predicts five. The only wrongly predicted note (2-nd note) of the AV-SVT system is also close to the ground truth since the MIDI note number difference is 1. To further explain the predictions of our AV-SVT system, we also visualize some frames of lip movements. Firstly, we display three frames $\bm{x}_{t_1-1}^V, \bm{x}_{t_1}^V, \bm{x}_{t_1+1}^V$ located at the onset $t_1$ of the 3-rd note. We notice that from $t_1-1$ to $t_1+1$, the subject slightly opened her mouth, which marks the transition from silence to 3-rd note. Moreover, we also display three frames $\bm{x}_{t_2-1}^V, \bm{x}_{t_2}^V, \bm{x}_{t_2+1}^V$ located at the offset $t_2$ of the last note. It is observed that from $t_2-1$ to $t_2+1$, the subject gradually closed her mouth. As displayed in Fig. \ref{fig_qual}(b), we also visualize the predictions of our A-SVT and AV-SVT systems in an environment with 0 dB musical accompaniment. Fig. \ref{fig_qual}(a) and (b) further validate that our AV-SVT system captures the transitions of consecutive note events and improves the noise robustness compared to audio-only systems.

\subsection{Ablation Study}
\subsubsection{Effectiveness of our adaptation strategy}

\begin{table}[t]
\caption{COnPOff / COnP / COn / COff F1-score (\%) of our SVT system on different test sets. We compare the performances of our SVT system when using different adaptation strategies.}
	\centering
	\begin{tabular}{l|l|c|c|c}
		\toprule
		Dataset & Metric & Ours & \multicolumn{2}{c}{Baseline} \\
        &(\%) $\uparrow$& & variant 1 & variant 2\\
		\midrule
  \multirow{4}{4.5em}{\textbf{N20EMv2 valid}} & COnPOff & \textbf{61.83} & 56.89 ({\color{purple}-\,\,\,4.94}) & 59.24 ({\color{purple}-\,\,\,2.59})\\
		& COnP & \textbf{68.42} & 63.39 ({\color{purple}-\,\,\,5.03}) & 65.99 ({\color{purple}-\,\,\,2.43})\\
		& COn & \textbf{92.18} & 91.50 ({\color{purple}-\,\,\,0.68}) & 91.17 ({\color{purple}-\,\,\,1.01}) \\
        & COff & \textbf{89.80} & 89.09 ({\color{purple}-\,\,\,0.71}) & 89.62 ({\color{purple}-\,\,\,0.18})\\
		\midrule
		\multirow{4}{4.5em}{\textbf{N20EMv2 test}} & COnPOff & \textbf{73.06} & 70.16 ({\color{purple}-\,\,\,2.90}) & 69.90 ({\color{purple}-\,\,\,3.16})\\
		& COnP & \textbf{79.56} & 77.25 ({\color{purple}-\,\,\,2.31}) & 76.84 ({\color{purple}-\,\,\,2.72})\\
		& COn & \textbf{93.66} & 93.08 ({\color{purple}-\,\,\,0.58}) & 92.71 ({\color{purple}-\,\,\,0.95})\\
        & COff & \textbf{91.78} & 91.22 ({\color{purple}-\,\,\,0.56}) & 91.21 ({\color{purple}-\,\,\,0.57})\\
        \midrule
		\multirow{3}{4.5em}{\textbf{MIR-ST500}} & COnPOff & \textbf{52.84} & 50.78 ({\color{purple}-\,\,\,2.06}) & 51.43 ({\color{purple}-\,\,\,1.41})\\
		& COnP & \textbf{70.00} & 68.75 ({\color{purple}-\,\,\,1.25}) & 68.89 ({\color{purple}-\,\,\,1.11})\\
		& COn & \textbf{78.05} & 77.23 ({\color{purple}-\,\,\,0.82}) & 77.98 ({\color{purple}-\,\,\,0.07})\\
		\midrule
		\multirow{3}{4.5em}{\textbf{TONAS}} & COnPOff & \textbf{24.08} & 21.63 ({\color{purple}-\,\,\,2.45}) & 22.55 ({\color{purple}-\,\,\,1.53})\\
		& COnP & \textbf{36.87} & 34.60 ({\color{purple}-\,\,\,2.27}) & 36.72 ({\color{purple}-\,\,\,0.15})\\
		& COn & \textbf{64.38} & 63.01 ({\color{purple}-\,\,\,1.37}) & 63.48 ({\color{purple}-\,\,\,0.90})\\
		\midrule
		\multirow{3}{4.5em}{\textbf{ISMIR2014}} & COnPOff & \textbf{62.42} & 61.03 ({\color{purple}-\,\,\,1.39}) & 57.97 ({\color{purple}-\,\,\,4.45})\\
		& COnP & \textbf{75.91} & 74.25 ({\color{purple}-\,\,\,1.66}) & 72.21 ({\color{purple}-\,\,\,3.70})\\
		& COn & \textbf{93.02} & 91.84 ({\color{purple}-\,\,\,1.18}) & 92.16 ({\color{purple}-\,\,\,0.86})\\
		\bottomrule
	\end{tabular}
    
    \label{tbl-exp-E1}
\end{table}

To adapt the self-supervised learning (SSL) models from the speech domain to the SVT tasks, we propose skipping the finetuning of SSL models on the ASR task and directly finetuning SSL models on the SVT task in a linear-probing and full-finetuning (LP-FT) fashion. To prove the effectiveness of this adaptation strategy, we conduct an ablation study on our A-SVT system by creating two variants. For ``variant 1", we keep the finetuning of SSL models on the ASR task, followed by full-finetuning for ten epochs on the SVT task (the order of $(a)\rightarrow(b)\rightarrow(d)$ in Fig. \ref{fig_lp_ft}). For ``variant 2", we skip the finetuning of SSL models on the ASR task, followed by full-finetuning on the SVT task for ten epochs (the order of $(a)\rightarrow(d)$ in Fig. \ref{fig_lp_ft}. 

As displayed in Table \ref{tbl-exp-E1}, our A-SVT system using the proposed adaptation strategy consistently outperforms the two variants on all datasets in terms of all metrics, including ID testing (MIR-ST500, N20EMv2 valid/test set) and OOD testing (TONAS, ISMIR2014). The results indicate the superiority of our adaptation strategy.

\subsubsection{Ablation on model architecture choice}
\begin{table}[t]
\caption{COnPOff / COnP / COn / COff F1-score (\%) of AV-HuBERT-based and wav2vec 2.0-based SVT systems on different test sets.}
	\centering
	\begin{tabular}{l|l|c|c|c}
		\toprule
		Dataset & Metric & Ours & \multicolumn{2}{c}{AV-HuBERT} \\
        & (\%) $\uparrow$ & & 25 Hz & 50 Hz \\
		\midrule
  \multirow{4}{4.5em}{\textbf{N20EMv2 valid}} & COnPOff & \textbf{61.83} & 45.50 ({\color{purple}-16.33}) & 22.25 ({\color{purple}-39.58})\\
		& COnP & \textbf{68.42} & 52.22 ({\color{purple}-16.20}) & 29.37 ({\color{purple}-39.05})\\
		& COn & \textbf{92.18} & 86.67  ({\color{purple}-\,\,\,5.51}) & 80.12 ({\color{purple}-12.06})\\
        & COff & \textbf{89.80} & 87.23 ({\color{purple}-\,\,\,2.57}) & 74.57 ({\color{purple}-15.23})\\
		\midrule
		\multirow{4}{4.5em}{\textbf{N20EMv2 test}} & COnPOff & \textbf{73.06} & 62.96 ({\color{purple}-10.10}) & 33.37 ({\color{purple}-39.69})\\
		& COnP & \textbf{79.56} & 70.32 ({\color{purple}-\,\,\,9.24}) & 44.60 ({\color{purple}-34.96})\\
		& COn & \textbf{93.66} & 90.63 ({\color{purple}-\,\,\,3.03}) & 81.25 ({\color{purple}-12.41})\\
        & COff & \textbf{91.78} & 90.77 ({\color{purple}-\,\,\,1.01}) & 77.21 ({\color{purple}-14.57})\\
        \midrule
		\multirow{3}{4.5em}{\textbf{MIR-ST500}} & COnPOff & \textbf{52.84} & 42.95 ({\color{purple}-\,\,\,9.89}) & 21.13 ({\color{purple}-31.71})\\
		& COnP & \textbf{70.00} & 61.06 ({\color{purple}-\,\,\,8.94})  & 35.36 ({\color{purple}-34.64})\\
		& COn & \textbf{78.05} & 74.03 ({\color{purple}-\,\,\,4.02}) & 66.26 ({\color{purple}-11.79})\\
		\midrule
		\multirow{3}{4.5em}{\textbf{TONAS}} & COnPOff & \textbf{24.08} & 19.84 ({\color{purple}-\,\,\,4.24}) & \,\,\,5.22 ({\color{purple}-18.86})\\
		& COnP & \textbf{36.87} & 32.13 ({\color{purple}-\,\,\,4.74}) & 15.17 ({\color{purple}-21.70})\\
		& COn & \textbf{64.38} & 62.62 ({\color{purple}-\,\,\,1.76}) & 43.33 ({\color{purple}-21.05})\\
		\midrule
		\multirow{3}{4.5em}{\textbf{ISMIR2014}} & COnPOff & \textbf{62.42} & 46.92 ({\color{purple}-15.50}) & 25.40 ({\color{purple}-37.02})\\
		& COnP & \textbf{75.91} & 56.35 ({\color{purple}-19.56}) & 35.89 ({\color{purple}-40.02})\\
		& COn & \textbf{93.02} & 86.43 ({\color{purple}-\,\,\,6.59}) & 81.76 ({\color{purple}-11.26})\\
		\bottomrule
	\end{tabular}
    
    \label{tbl-exp-E2}
\end{table}

As described in Section \ref{avhubert}, AV-HuBERT can accept both audio and video modality. Therefore, it is possible to build A-SVT / AV-SVT systems using AV-HuBERT as the audio-specific feature encoder. However, in our preliminary experiments, it is difficult for AV-HuBERT to learn powerful acoustic representations for the SVT tasks. To prove this, we build an A-SVT system using AV-HuBERT. We disable the video branch of AV-HuBERT. Unlike wav2vec 2.0, which accepts the raw waveform as input, AV-HuBERT requires audio signals' log filterbank energy feature as input. Specifically, the 26-dimensional features are extracted from the raw waveform at a stride of 10 ms. The four neighboring acoustic frames are then stacked together, resulting in a frame rate of 25 Hz \cite{shi2022avsr:avhubert}.

From Table \ref{tbl-exp-E2}, the performance of the A-SVT system based on AV-HuBERT is much worse than that based on wav2vec 2.0, especially for pitch estimation. We attribute the reason to the low frame rate of input to AV-HuBERT. Each frame of the feature extracted by AV-HuBERT is 40 ms, while the frame length of the wav2vec 2.0 feature is about 20 ms. Frame resolution is a significant factor in the accuracy of the A-SVT system. Therefore, we assume low frame resolution is one of the reasons why the performance of the AV-HuBERT-based A-SVT system is inferior to that of the wav2vec 2.0-based A-SVT system. We attempt to change the input frame rate to 50 Hz by stacking only two neighboring acoustic frames. Since the input dimension is also altered, we add another linear layer to adjust the dimensions. However, the performances become much worse than before. The reason is that AV-HuBERT was pretrained on acoustic features with a frame rate of 25 Hz. Modifying the input and model structure during the finetuning will drastically deteriorate the adaptation of SSL models to downstream tasks. 

\subsubsection{Discussion on adaptation of SSL models}

SSL has emerged as a paradigm in the recent deep learning community \cite{devlin2018bert, baevski2020wav2vec, he2020momentum, brown2020language, baevski2022data2vec}. One key aspect is the excellence of SSL models in downstream tasks, even in low-resource, few-shot, zero-shot setups. We agree with \cite{kumar2022fine} that pretrained features play an essential role. Therefore, during the finetuning, the distortion of pretrained features will result in performance drops on downstream tasks. From this perspective, modifying the input or model structure can distort the pretrained features of SSL models. In contrast, skipping the finetuning on the speech recognition tasks, linear probing before full finetuning, and adopting smaller learning rates for SSL models will preserve the pretrained features. 

\section{Conclusion}
In this work, we proposed an audio-visual singing voice transcription (AV-SVT) system based on self-supervised learning (SSL) models. We curated the first multimodal SVT dataset, N20EMv2, to implement our systems. We then proposed a new approach to adapt SSL models from the speech domain to the SVT tasks to mitigate the challenge of label insufficiency. Based on this, our audio-only SVT system outperformed state-of-the-art technologies significantly and generalized to out-of-domain singing data of different languages and styles. We then initialized the video-only SVT task and our system successfully detected about 80\% onset and offset of notes. Our ablation studies demonstrate the effectiveness of our adaptation strategy and model choices. Finally, our audio-visual SVT system showed excellence in clean and noisy scenarios through the experiments. Hence, our attempts validated that introducing additional modality can improve the noise robustness of the SVT system compared to the audio-only systems.

\bibliographystyle{IEEEtran}
\bibliography{reference}
\clearpage
\section{Additional details of datasets}
In this section, we includes more details about how we curate the N20EMv2 dataset and more comparisons between the N20EMv2 dataset and benchmark singing voice transcription (SVT) datasets. 

\subsubsection{Curation of our N20EMv2 dataset}
In the main paper, we mention that two music experts accomplished the annotation process of the N20EMv2 dataset. To ensure inter-rater reliability, we set several rules as guidelines before annotation, which are important for ensuring a higher quality of the N20EMv2 dataset. Firstly, notes are segmented according to both pitches and syllables. Different syllables are always considered as separate notes, while detailed criteria about the onset/offset/pitch labeling are as follows:
\begin{itemize}
    \item \textbf{Pitch}: Pitches with a duration longer than a semiquaver are considered individual notes as perceived by the annotators, while ornaments such as pitch bending at the beginning of the note or vibratos are not considered independent notes. The pitch of each note is annotated in semitonal resolution.
    \item \textbf{Onset}: The onset time of each note is marked as the start of the vowel in each syllable. If a syllable begins with a non-vowel sonorant, the annotators deliberately determine when the vowel is pronounced as onset. For instance, if the lyrics of a note is ``last" [la:st], the onset is placed at the beginning of ``a" [a:] instead of ``l" [l].
    \item \textbf{Offset}: The offset time of each note is marked when there are no significant patterns in the audio spectrogram or the next note starts.
\end{itemize}
After the initial annotation, the two experts scrutinize each other's labeling results to reach final agreements.

\subsubsection{Comparison with the benchmark SVT datasets}
\begin{table*}[b]
\caption{Comparisons among various datasets for the SVT task.}
	\centering
	\begin{tabular}{l|c|c|c|c|c|c|c|c|c}
		\toprule
		Dataset & Size & Ave length & \makecell{Total \\ length} & Year & Multimodal & Singing voice & Genre & Method & Pitch resolution \\
		\midrule
		\textbf{TONAS} & \makecell{72 songs, \\ 2983 notes} & 30 s & 36 min & 2013 & \textbf{No} & Flamenco expert & Flamenco songs & \makecell{Labeled by \\ experts} & Semitone \\
        \midrule
		\textbf{ISMIR2014} & \makecell{38 songs, \\ 2153 notes} & 15 to 86 s & 19 min & 2014 & \textbf{No} & \makecell{14 children, \\ 13 adult male, \\ 11 adult female} & \makecell{Pop songs, \\children songs} & \makecell{Labeled by \\ experts} & Cent \\
        \midrule
        \textbf{MIR-ST500} & \makecell{500 songs, \\ \textgreater 160k notes} & 3 to 5 min & $\approx$ 30 h & 2021 & \textbf{No} & \makecell{Online \\ published songs} & \makecell{Pop songs \\ in Chinese} & \makecell{Labeled by \\ non-experts} & Semitone \\
		\midrule
        \textbf{\makecell{N20EMv2 \\ (Ours)}} & \makecell{157 songs, \\ 38857 notes} & 2 to 5 min & $\approx$ 8.4 h & 2022 & \textbf{Yes} & \makecell{Recorded songs \\ from amateurs} & \makecell{Pop songs \\ in English} & \makecell{Labeled by \\ experts} & Semitone \\
		\bottomrule
	\end{tabular}
    \label{tbl-datasets_cmp}
\end{table*}

In Table \ref{tbl-datasets_cmp}, we compare our curated N20EMv2 dataset with three benchmark SVT datasets including TONAS \cite{gomez2013towards}, ISMIR2014 \cite{molina2014evaluation}, and MIR-ST500 \cite{wang2021preparation}. Firstly, our N20EMv2 is the first multimodal SVT dataset. Secondly, N20EMv2 is much more large-scale than TONAS / ISMIR2014 and has more accurate annotations than MIR-ST500. It is noticed that except for ISMIR2014 which labeled the pitch in cents resolution, all the other datasets labeled the pitch as semitones. Because almost all modern music is based on a 12-tonal equal temperament scale, we follow the convention of labeling the pitch in semitones to balance the labeling precision and efficiency.

\section{Additional experiments}
Besides the experiments conducted in the main paper, we also run two additional experiments to (a) evaluate the annotation quality of our curated dataset N20EMv2; (b) explore the effects of video input frame rate on the performance of our video-only SVT system.

\subsubsection{Evaluation of N20EMv2 Annotation Quality}
We evaluate the annotation quality of our N20EMv2 by comparing it to the MIR-ST500 dataset. Specifically, we train our audio-only SVT systems separately using the MIR-ST500 training set and N20EMv2 training set. The former is marked as ``Ours variant 1" in the main paper. Afterward, we conduct out-of-domain (OOD) testing, reflecting the generalization abilities of two trained SVT systems on singing data from unseen domains. We select TONAS and ISMIR2014, two datasets for OOD testing.

As shown in Table \ref{tbl-quality}, it is noticed that the SVT system trained on the N20EMv2 training set demonstrates better OOD testing performance even though the number of songs in the MIR-ST500 training set (400 songs) is larger than that in our N20EMv2 training set (123 songs). Training on our N20EMv2 dataset improves the performance of our audio-only SVT system on TONAS by $5.44\%$ in COnPOff, $3.47\%$ in COnP, $10.51\%$ in COn. While on the ISMIR2014 dataset, two SVT systems perform similarly for onset detection. However, the SVT system trained on N20EMv2 shows performance gains in COnPOff and COnP. Through this experiment, we conclude that the annotation quality of our N20EMv2 dataset is better than MIR-ST500.

\begin{table}
\caption{COnPOff / COnP / COn F1-score (\%) of our A-SVT system on the TONAS and ISMIR2014 datasets. We compare the performance of our SVT system when using different training sets.}
	\centering
	\begin{tabular}{l|l|c|c}
		\toprule
		Dataset & Metric (\%) $\uparrow$ & MIR-ST500 & N20EMv2 \\
		\midrule
		\multirow{3}{6em}{\textbf{TONAS}} & COnPOff &  12.71  & \textbf{18.15} (\textbf{{\color{teal}+\,\,\,5.44}}) \\
		& COnP & 25.24  & \textbf{28.71} (\textbf{{\color{teal}+\,\,\,3.47}}) \\
		& COn & 52.77  & \textbf{63.28} (\textbf{{\color{teal}+10.51}}) \\
		\midrule
		\multirow{3}{6em}{\textbf{ISMIR2014}} & COnPOff &  52.36  & \textbf{59.46} (\textbf{{\color{teal}+\,\,\,7.10}}) \\
		& COnP & 70.38  & \textbf{73.19} (\textbf{{\color{teal}+\,\,\,2.81}}) \\
		& COn & \textbf{92.77}  & 91.88 (\textbf{{\color{purple}-\,\,\,0.89}}) \\
		\bottomrule
	\end{tabular}
    
    \label{tbl-quality}
\end{table}

\subsubsection{Ablation on video frame rate choice}
As mentioned in the main paper, AV-HuBERT was pretrained on video of lip movements at a frame rate of 25 Hz \cite{shi2022avsr:avhubert}. However, the frame rate of acoustic features extracted by wav2vec 2.0 is 50 Hz \cite{baevski2020wav2vec}. To facilitate the feature fusion of audio and video modality and increase the frame resolution, we decide to use videos at a frame rate of 50 Hz. Unlike changing the frame length of acoustic features extracted by AV-HuBERT, there is no need to modify its model structure when changing the input video frame rate. Therefore, we conduct this ablation study to validate our choice of video frame rate by comparing our video-only SVT systems accepting the videos in different frame rates. We keep other training configurations the same for fair comparisons.

As shown in Table \ref{tbl-exp-E3}, the V-SVT system with 50 Hz video input performs better than that with 25 Hz video input, especially for onset detection. The improvement is not as significant as we expect. We attribute this phenomenon to that changing the video input frame rate distorts the pretrained features, which neutralizes the improvements brought by higher frame resolution. Considering the feature fusion and the slight performance gains, we set the frame rate of video input to AV-HuBERT as 50 Hz. 

\begin{table}[t]
\caption{COnPOff / COnP / COn / COff F1-score (\%) of our V-SVT system on the N20EMv2 valid/test set. We compare the performances using different video input frame rates (Hz).}
	\centering
	\begin{tabular}{l|l|c|c}
		\toprule
		Dataset & Metric (\%) $\uparrow$ & 25 Hz & 50 Hz \\
		\midrule
		\multirow{4}{5em}{\textbf{N20EMv2 valid}} & COnPOff &  \,\,4.14 & \,\,4.45 (\textbf{{\color{teal}+\,\,\,0.31}})\\
		& COnP &  \,\,5.38 & \,\,6.16 (\textbf{{\color{teal}+\,\,\,0.78}})\\
		& COn & 74.89 & 77.14 (\textbf{{\color{teal}+\,\,\,2.25}})\\ 
		& COff & 76.01 & 74.68 (\textbf{{\color{purple}-\,\,\,1.33}})\\
		\midrule
		\multirow{4}{5em}{\textbf{N20EMv2 test}} & COnPOff & \,\,5.38 & \,\,6.84 (\textbf{{\color{teal}+\,\,\,1.46}})\\
		& COnP & \,\,7.31 & \,\,8.79 (\textbf{{\color{teal}+\,\,\,1.48}})\\
		& COn & 76.40 & 78.62 (\textbf{{\color{teal}+\,\,\,2.22}})\\
		& COff & 78.77 & 78.83 (\textbf{{\color{teal}+\,\,\,0.06}})\\
		\bottomrule
	\end{tabular}
    
    \label{tbl-exp-E3}
\end{table}
\section{Additional visualization results}
\begin{figure*}[t!]
\centering
\includegraphics[width=\linewidth]{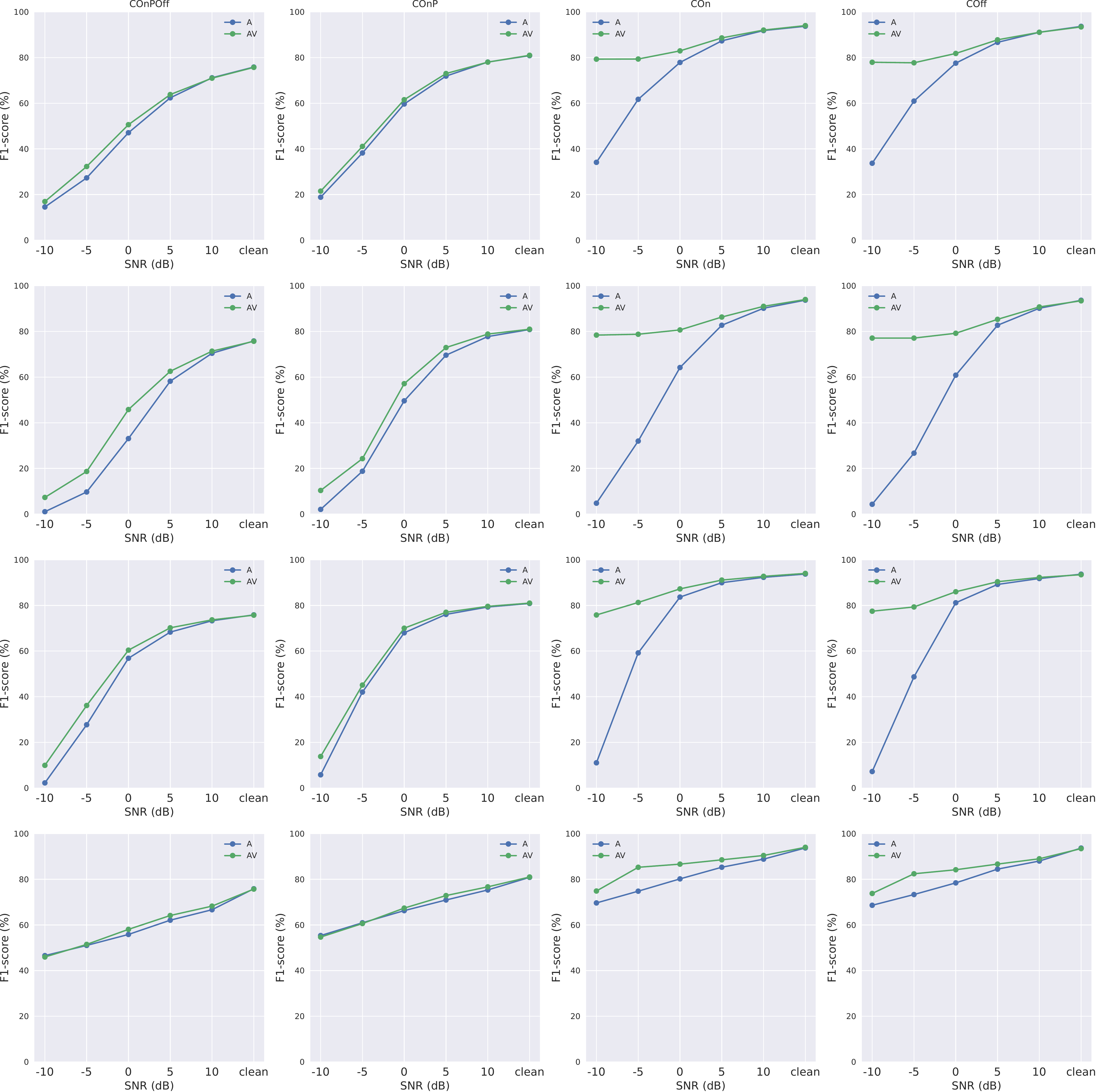}
\caption{Complete quantitative comparison of audio-only SVT system and audio-visual SVT system on N20EMv2 test set. For each row, we compare the COnPOff, COnP, COn, COff four metrics of two SVT systems under the same noise perturbations. From the top to bottom, we use the musical accompaniment, babble noise, white noise, natural noise as the perturbations, respectively.}
\label{fig_noise_complete}
\end{figure*}

We visualize the complete quantitative comparison of the audio-only and audio-visual SVT systems on the N20EMv2 test set, including COnPOff, COnP, COn, COff four metrics in Fig. \ref{fig_noise_complete}.

\end{document}